\documentclass{article}

\usepackage[final, nonatbib]{neurips_2025_ml4ps} 
\usepackage[numbers,sort&compress]{natbib}

\usepackage[utf8]{inputenc} 
\usepackage[T1]{fontenc}    
\usepackage{hyperref}       
\usepackage{url}            
\usepackage{booktabs}       
\usepackage{amsfonts}       
\usepackage{nicefrac}       
\usepackage{microtype}      
\usepackage{xcolor}         
\usepackage{amsmath}
\usepackage{graphicx}

\usepackage[acronym]{glossaries}
\glsdisablehyper
\makeglossaries

\newcommand{\mrm}[1]{\mathrm{#1}}
\newcommand{\mbf}[1]{\mathbf{#1}}
\newcommand{\mcal}[1]{\mathcal{#1}}

\newcommand{\mbs}[1]{\boldsymbol{#1}}

\newcommand{\PhiAB}{\phi_{\mrm{AB}}}
\newcommand{\PhiNN}{\tilde{\phi}_{\mrm{NN}}}
\newcommand{\PhiNNO}{\tilde{\phi}_{\mrm{NN}}{}_0}

\newacronym{NODE}{NODE}{Neural Ordinary Differential Equation}
\newacronym[plural=BFEs, firstplural=basis function expansions (BFEs)]{BFE}{BFE}{basis function expansion}
\newacronym[plural=PINNs, firstplural=physics-informed neural networks (PINNs)]{PINN}{PINN}{physics-informed neural network}

\title{Physics-Informed Neural Networks for Modeling Galactic Gravitational Potentials}

\author{
  Charlotte Myers\textsuperscript{1} \\
  \texttt{c\_myers@mit.edu} \\
  \And
  Nathaniel Starkman\textsuperscript{1,2,3} \\
  \texttt{starkman@mit.edu} \\
  \And
  Lina Necib\textsuperscript{1, 4} \\
  \texttt{lnecib@mit.edu} \\
  \AND
  \textsuperscript{1}MIT Department of Physics and MIT Kavli Institute for Astrophysics and Space Research, \\Massachusetts Institute of Technology, Cambridge, MA 02139, USA \\
  \textsuperscript{2}Brinson Prize Fellow \\
  \textsuperscript{3}Visiting Scholar, Case Western Reserve University \\
  \textsuperscript{4}The NSF AI Institute for Artificial Intelligence and Fundamental Interactions, \\
  Massachusetts Institute of Technology, Cambridge, MA 02139, USA
}

\begin{document}

\maketitle

\begin{abstract}
    We introduce a physics-informed neural framework for modeling static and time-dependent galactic gravitational potentials. %
    The method combines data-driven learning with embedded physical constraints to capture complex, small-scale features while preserving global physical consistency. %
    We quantify predictive uncertainty through a Bayesian framework, and model time evolution using a neural ODE approach. %
    Applied to mock systems of varying complexity, the model achieves reconstruction errors at the sub-percent level ($0.14\%$ mean acceleration error) and improves dynamical consistency compared to analytic baselines. %
    This method complements existing analytic methods, enabling physics-informed baseline potentials to be combined with neural residual fields to achieve both interpretable and accurate potential models.
\end{abstract}

\section{Introduction}\label{sec:introduction}

    Modeling the gravitational potential of a galaxy is a central problem in astrophysics, providing the link between the observed motions of stars and gas to the underlying distribution of mass \cite{Binney}. %
    Since most of this mass is in the form of dark matter -- interacting only via its gravitational influence -- potential models are among the few probes of its distribution and properties, while also serving as a foundation for modeling a galaxy's small-scale structure and dynamical evolution \citep{refId0, STRIGARI20131}. %

    Existing gravity modeling methods span a trade-off between interpretability, flexibility, and efficiency. %
    Analytic models employ interpretable, physically motivated forms (e.g. NFW halos), but struggle to model substructures and deviations from the idealized form \citep{Navarro_1997}. %
    \Glspl{BFE} trade for increased flexibility by representing the potential as a sum over modes. %
    However, for complex systems they can be inefficient due to requiring a large number of modes and misleading in their interpretation due to unphysical terms (e.g. negative mass components) and representation of noise by high-order terms \citep{BFEs}. %
    
    \Glspl{PINN} address many of these limitations by generalizing the concept of \glspl{BFE}, learning flexible and adaptive basis functions directly from data while embedding physical constraints into the training objective. %
    By constraining the hypothesis space to physically consistent functions, \glspl{PINN} reduce the required parameter count and improve generalization compared to purely empirical networks \citep{pinnsrecentadvances, RAISSI2019686}. %
    Recent advances, particularly the PINN-GM-III architecture, have achieved state-of-the-art performance in reconstructing gravitational fields of terrestrial objects with high accuracy and robust extrapolation \citep{pinngm}. %
    One compelling feature of the architecture is the combination of analytic baseline potentials with the neural networks, enabling e.g. a low-order BFE to be used in tandem with the generalizing network.
    
    In this work we make three main contributions: %
    (1) we extend the PINN-GM-III framework to model galactic potentials, where larger spatial scales, diverse structural components, and non-axisymmetric perturbations pose new challenges; %
    (2) we introduce a time-dependent neural ordinary differential equation (NODE) formulation that captures dynamical evolution; %
    (3) we incorporate Bayesian inference to quantify uncertainty in both static and evolving fields.

    \section{Model design}\label{sec:model_design}

        Our model combines analytic structure with learned residuals through six components: %
        (i) a loss that enforces the physical relation $\mbf{a}=-\nabla\phi$; %
        (ii) a coordinate transform that compactifies the spatial domain;
        (iii) radial scaling and (iv) analytic fusing, which factor out the dominant large–scale trends and reserve capacity for higher–order perturbations; %
        (v) a Bayesian treatment to capture epistemic uncertainty and to infer analytic scale parameters jointly with the residual field; %
        and (vi) a neural ODE formulation to constrain temporal evolution to a continuous, causal trajectory. %
        \subsection{Architecture}\label{sec:model_design:architecture}

            The model accepts \textit{N} paired vectors of positions and accelerations, $(\mbf{x}_i, \mbf{a}_i)$, sampled from the target gravitational field. %
            The network, parameterized by $\mbs{\theta}$, outputs the predicted potential $\phi(\mbf{x}_i)$, which is differentiated with respect to position to produce an acceleration vector. %
            The physics-informed loss $\mcal{L}$ enforces the constraint $\mbf{a} = -\nabla\phi$ by combining absolute and relative acceleration errors: %
            \begin{equation}\label{eq:loss}
                \mcal{L}(\mbs{\theta})
                    = \frac{1}{N} \sum_{i=1}^{N} \left( 
                        \left\| -\nabla \phi(\mbf{x}_i | \mbs{\theta}) - \mbf{a}_i \right\| 
                        + 
                        \lambda_\text{r} \frac{\left\| -\nabla \phi(\mbf{x}_i | \mbs{\theta}) - \mbf{a}_i \right\|}{\left\| \mbf{a}_i \right\|}
                      \right).
            \end{equation}
            The relative term, weighted by $\lambda_\text{r}$, prevents loss of accuracy at large radii where accelerations decay toward zero. %
            Because the loss is written in terms of the gradient of the learned potential, each optimization step explicitly penalizes violations of the force--potential relation $\mbf{a}=-\nabla\phi$, biasing the model capacity toward physically consistent solutions.

    \subsection{Input representation}\label{sec:model_design:input_representation}
        
        We adopt the 5D spherical coordinate system of \citet{pinngm}, which maps the domain to a finite range to improve numerical stability. %
        In this parameterization, the three angular coordinates $(s, t, u)$ specify direction, while two compactified radial coordinates $(r_i, r_e)$ map the interior and exterior regions to bounded intervals. %
        This construction keeps both $\phi$ and its spatial derivatives within a well-conditioned range across the domain, allowing the physics-informed loss in Eq.~\eqref{eq:loss} to weight violations of $\mbf{a}=-\nabla\phi$ consistently over all scales. %


    \subsection{Design features: radial scaling and analytic fusing}\label{sec:model_design:design_features}

        Expanding on the design of \citet{pinngm}, we factor out the analytic radial trend from the full potential $\phi(\mbf{x})$ using a scaling function $n(\mbf{x})$, and train the network to predict a scaled residual potential $\PhiNN(\mbf{x}) = \phi(\mbf{x}) \cdot n(\mbf{x})$. %
        This compresses the dynamic range of the output, improving numerical conditioning of both $\phi$ and $\nabla\phi$ and directing the loss in Eq.~\eqref{eq:loss} toward structure that deviates from the dominant radial fall-off. %
        
\begin{figure}
            \hspace{-0.8cm} \includegraphics[width=1.14\linewidth]{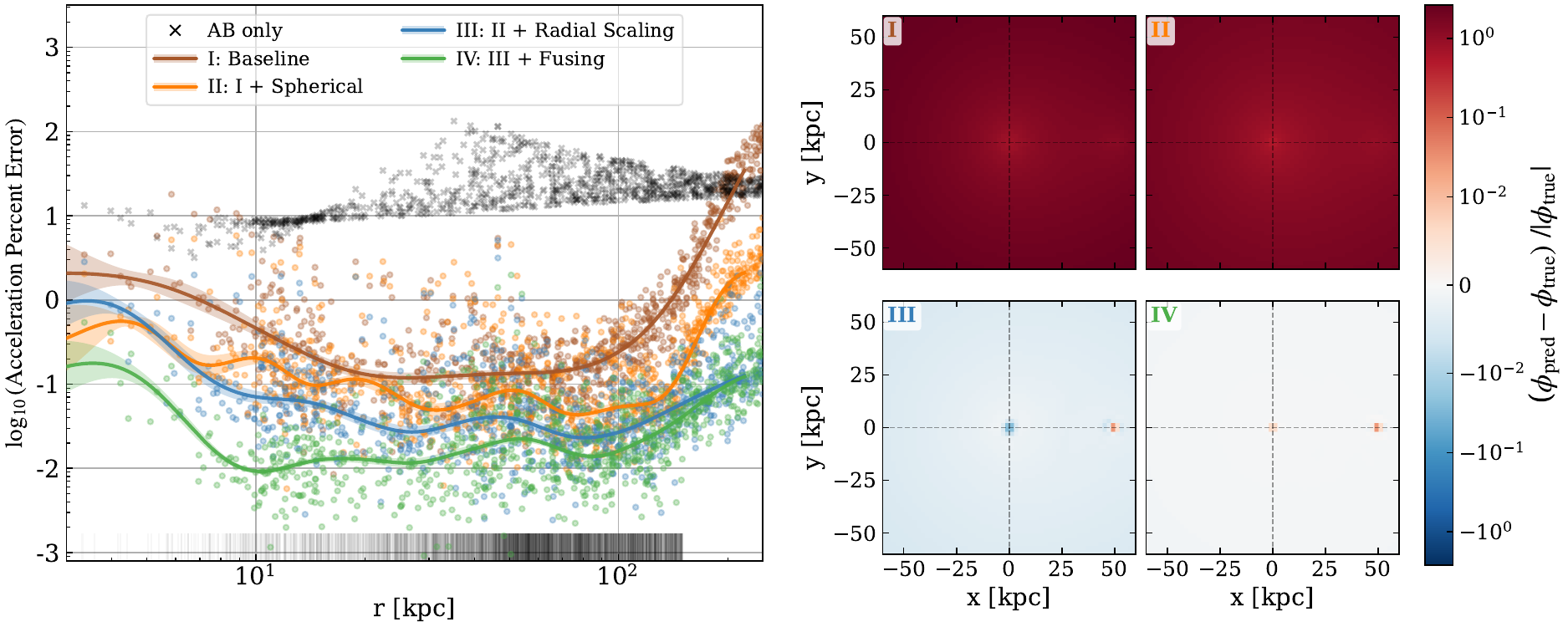}
            \caption{%
                \textbf{Impact of model design choices on reconstructing the MW--LMC system.}  %
                \textbf{Left}: Radial profile of relative acceleration percent error ($100 \cdot \tfrac{\|\mbf{a}_\text{pred} - \mbf{a}_\text{true}\|}{\|\mbf{a}_\text{true}\|}$). %
                Black dashes mark the training point locations, and black crosses represent the performance of an analytic MW model without NN correction. %
                \textbf{Right}: Relative residual of the reconstructed potential field in the $x$–$y$ plane.} %
            \label{fig:ablation}
        \end{figure}

        We also incorporate an analytic baseline (AB) model $\PhiAB(\mbf{x})$ that captures the known (generally large-scale, low-order) structure of the gravitational field. %
        Rather than relearning these well-understood features, the network predicts only the scaled residual component $\PhiNN(\mbf{x})$. %
        The full potential is reconstructed as %
        \begin{equation}\label{eq:phi_fused}
            \phi(\mbf{x}) = \PhiAB(\mbf{x}) + \PhiNN(\mbf{x}|\mbs{\theta}) / n(\mbf{x}) \, .
        \end{equation}
        This fusion leverages the interpretability and reliability of analytic modeling while reserving the network's flexibility for capturing higher-order perturbations. %
        Setting \(\PhiAB\!\equiv\!0\) recovers the scaling-only variant. %
        In Figure \ref{fig:ablation}, we show the effect of these design features on the reconstruction of the acceleration and potential fields of the Milky Way-LMC system (see Section \ref{sec:performance} for details). %

    \subsection{Bayesian framework}\label{sec:model_design:design_framework}
    
        To quantify epistemic uncertainty, we adopt a Bayesian neural network (BNN) framework.
        Unlike conventional neural networks, which produce fixed parameter values, a BNN treats each network weight and bias as a random variable with an associated probability distribution \cite{bnn2}.
        We implement this framework in \texttt{NumPyro} and approximate posterior distributions using stochastic variational inference (SVI).
        
        We place truncated-normal priors on the analytic parameters $\mbs{\alpha}$, constrained to physically plausible ranges, and zero-mean Gaussian priors with variance $\sigma_\theta^2$ on the neural network weights $\mbs{\theta}$.
        Rather than specifying an explicit generative likelihood for the acceleration data, we incorporate the physics-informed acceleration loss (\autoref{eq:loss}) directly into the variational objective as an unnormalized log-density term.
        The variational posterior $q(\mbs{\alpha},\mbs{\theta})$ is taken to be a diagonal Gaussian guide (\texttt{AutoNormal}) and is optimized by maximizing this augmented ELBO.
        To stabilize joint inference of analytic and residual components, training proceeds in two stages: an initial phase with a narrow neural weight prior ($\sigma_\theta=0.01$), encouraging the analytic baseline to dominate, followed by a phase with relaxed neural priors ($\sigma_\theta=1.0$) that allows the network to capture residual structure. 
        Each stage is initialized from the previous variational solution to ensure continuity in the posterior evolution. %
    
        \begin{figure}[t]
            \hspace{-1.5cm} 
            \includegraphics[width=1.11\linewidth]{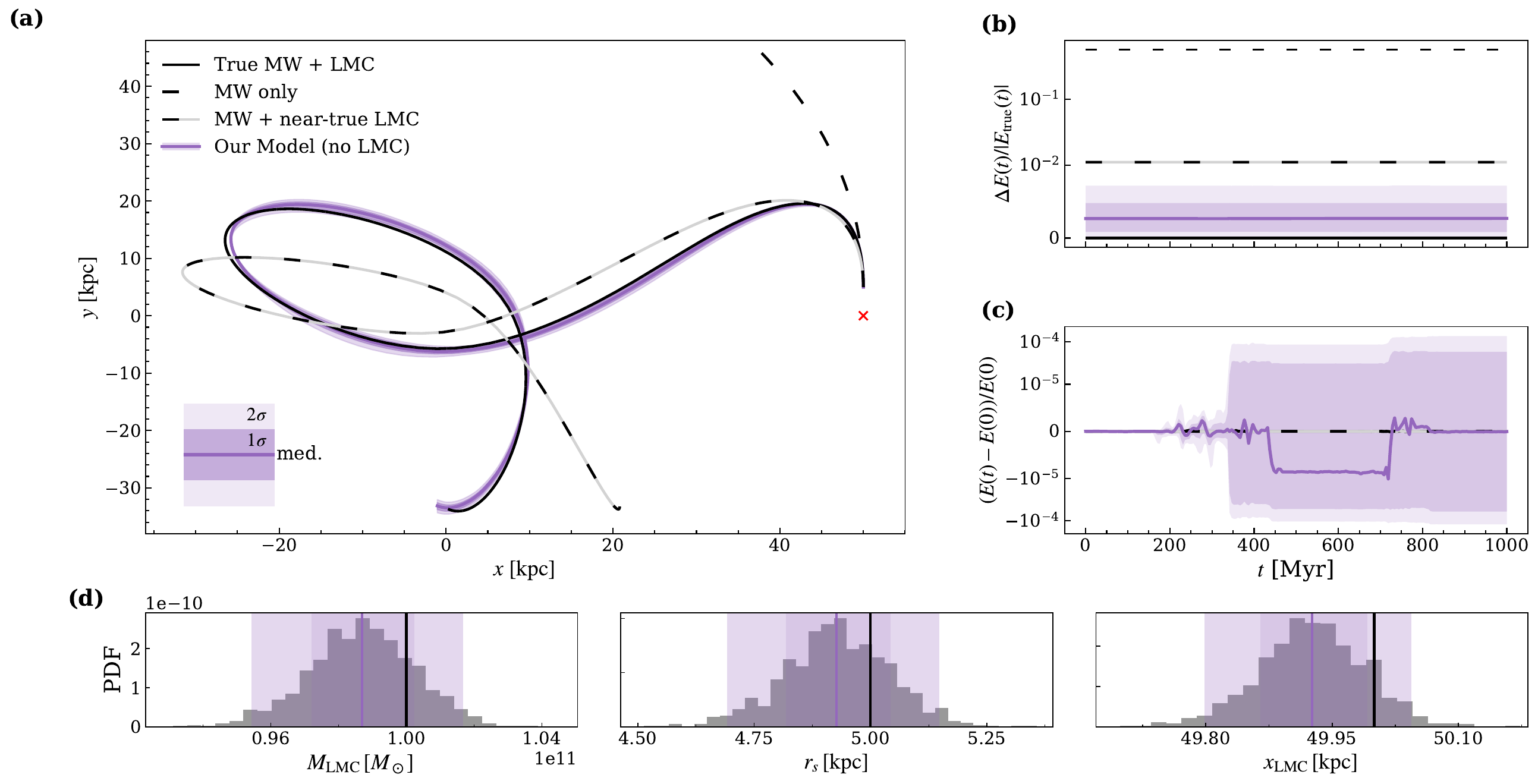}
            \caption{%
                \textbf{Model performance on the MW--LMC test system} %
                \textbf{(a)} Orbit initialized at the LMC center with the local circular velocity and integrated for 
            $1\,\mrm{Gyr}$ under four models: the true potential, an analytic MW--only baseline without the LMC, an MW-LMC model with misspecified parameters (LMC center offset by $1\,\mrm{kpc}$ and scale radius $r_s$ misestimated by $2\%$), and the BNN reconstruction. %
                \textbf{(b)} Relative deviation from true energy, where $\Delta E(t) = E_{\text{pred}} - E_{\text{true}} $. %
                \textbf{(c)} Relative energy drift along the predicted orbit.
                \textbf{(d)} Posterior distributions of the inferred LMC parameters (mass, scale radius, and Galactocentric distance).
                Each posterior sample is obtained by reconstructing the full learned potential and fitting it to a MW-LMC model with the MW parameters held fixed to their true values.
                Uncertainty envelopes in all panels correspond to the 16--84th ($1\sigma$) and 2.5--97.5th ($2\sigma$) percentiles across 1000 posterior draws.
                }
    
            \label{fig:accel_orbit_errs}
        \end{figure}

        \begin{figure}
            \centering
            \includegraphics[width=0.95\linewidth]{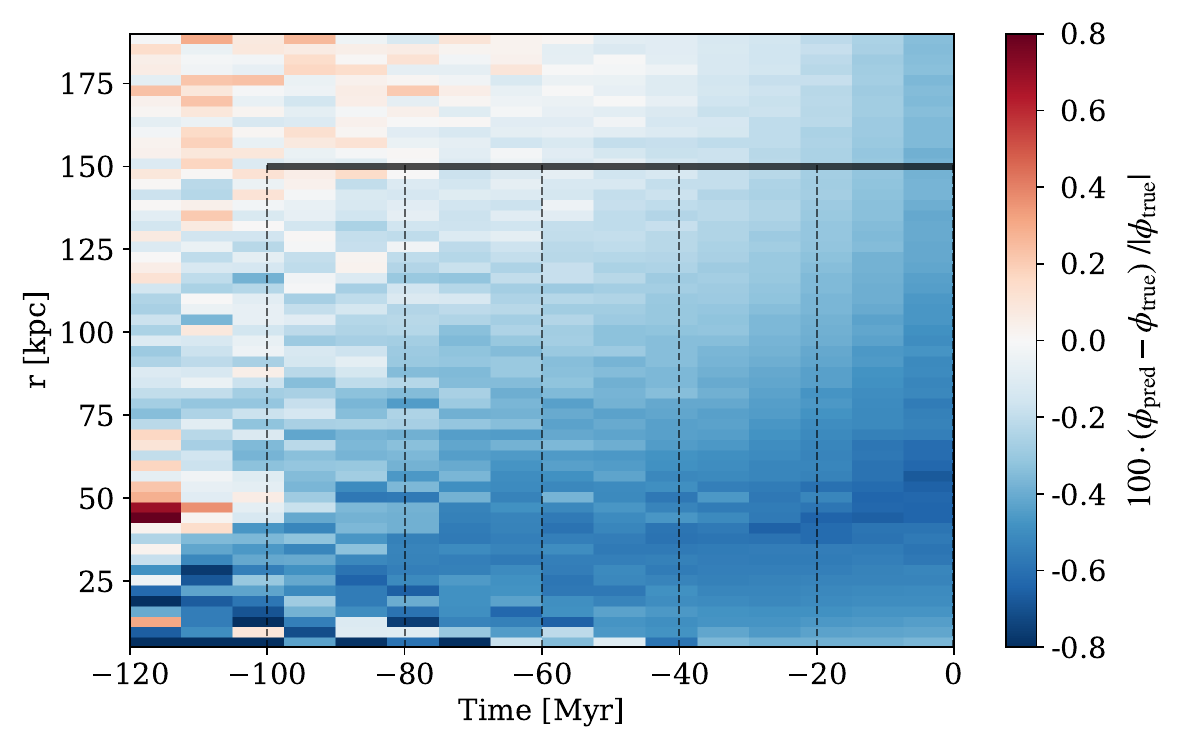}
            \caption{%
                \textbf{Relative percent error of the time-dependent MW--LMC potential reconstruction}, binned by radius and time relative to the present ($t=0$; $t<0$ earlier). %
                The dashed vertical lines mark training snapshot times; the solid horizontal line indicates the maximum training radius.%
            }
            \label{fig:potential_evolution}
        \end{figure}
    
    \subsection{Modeling time dependence}\label{sec:model_design:modeling_time_dependence}

        We adopt a NODE formulation in which the network predicts the time derivative $f_{\text{NN}} \equiv \frac{\mrm{d}}{\mrm{d}t}\PhiNN$ (parameterized by $\mbs{\theta}_2$) of the scaled residual potential field and the initial spatial correction (parameterized by $\mbs{\theta}_1$). %
        The full potential is then reconstructed as: %
        \begin{equation}\label{eq:node}
            \phi(\mbf{x}, t)
                =   \PhiAB(\mbf{x}, t)
                  + \frac{1}{n(\mbf{x}, t)} \cdot \ \left[ \PhiNN(\mbf{x},t|\mbs{\theta}) = \PhiNNO(\mbf{x}| \mbs{\theta}_1) + \int_{0}^{t} f_{\text{NN}}(\mbf{x}, t' | \mbs{\theta}_2)\ dt' \right],
        \end{equation}
        where $\PhiNNO(\mbf{x}| \mbs{\theta}_1)$ denotes the initial correction at $t=0$. %
        The time integral is evaluated numerically via Gauss–Legendre quadrature. %

        A naive approach to modeling temporal variation is to append time as a fourth coordinate in the network input. %
        However, this approach treats time as an unordered feature and does not enforce the causal structure of the evolution. %
        By contrast, the NODE approach constrains the evolution to follow a continuous trajectory, ensuring temporal consistency across the learned evolution. %


    \subsection{Implementation details}\label{sec:model_design:implementation}
        
\begin{figure}
            \centering
            \includegraphics[width=0.80\linewidth]{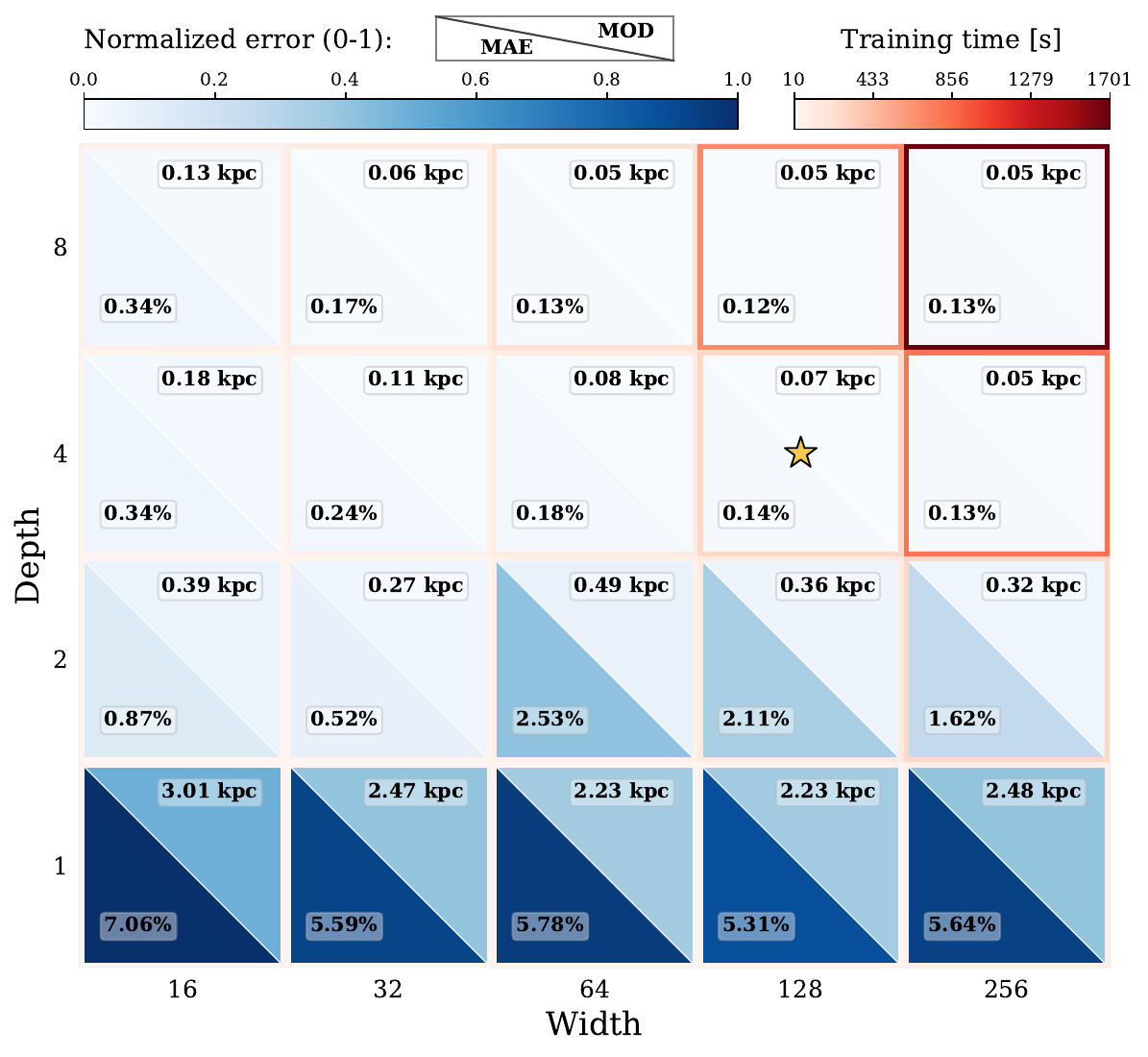}
            \caption{%
                \textbf{Effect of network size on performance.} %
                Each square corresponds to one network configuration, where the diagonal separates the two metrics: the lower-left shows the mean acceleration error (MAE, $\%$), evaluated on 32,768 testing points; the upper-right shows the mean orbit deviation (MOD, kpc), computed on a set of 100 orbits initialized at randomly sampled positions and integrated for 500 Myr. Training time is encoded by the color of each cell border. %
                The star marks the configuration selected for follow-up tests. %
            }
            \label{fig:depthwidth}
        \end{figure}

        We implement the network in \texttt{JAX} with \texttt{optax} for optimization; training uses the Adam optimizer with an initial learning rate of $3\times 10^{-3}$ and an exponential schedule that halves the rate every $1000$ epochs. 
        All experiments reported in \autoref{sec:performance} were run on a standard Apple M2 CPU; wall-clock training times are summarized alongside accuracy in \autoref{fig:depthwidth}. %
        Scripts for data generation, model training, and evaluation are available on
    \href{https://github.com/charlottemyers/galactoPINNs.git}{GitHub},
        with the corresponding datasets and trained models archived on
        \href{https://zenodo.org/records/18149747?preview=1&token=eyJhbGciOiJIUzUxMiJ9.eyJpZCI6IjdmYTZlOTk1LTEyYTYtNGJiNS1hNWY2LTUzMDY0OTc5OGM3MCIsImRhdGEiOnt9LCJyYW5kb20iOiIzOTA4OTdkNmRlNjEzMGEzZWZkZTM2Y2U4NjAwZTZmMiJ9.HTHewDij5ijl0_LdzEqX7hDx2oIHvMLMl7kq6CVdKGepSbEmieTCepoBEzdDgNfvWv3fusCBEpSOaaY61nr23w}{Zenodo}.
 
\section{Performance}\label{sec:performance}

    We assess the model through a series of controlled tests, the most challenging of which models the Large Magellanic Cloud (LMC) — the most massive satellite of the Milky Way (MW) — as a perturbation to an analytic baseline MW potential. %

    \textbf{Static field reconstruction:} The network is tasked with learning the LMC-induced perturbation as well as residual substructure omitted from the analytic baseline (e.g., the bulge and nucleus). %
    The global scale parameters of the host potential—the halo scale radius $r_s$ and the disk scale lengths $a$ and $b$—are jointly inferred within a Bayesian framework, as discussed in Section \ref{sec:model_design:design_framework}. %

    In this MW–LMC configuration (with the LMC placed 50 kpc from the Galactic center; \citep{10.1093/mnras/stab3726}), we train a 4$\times$128 dense network for 10,000 epochs using Adam optimization, and setting $\lambda_\text{r} = 0.1$. %
    The training set consists of 4096 samples drawn via density-based rejection sampling to reflect the underlying mass distribution. %
    To assess dynamical consistency, we integrate test-particle orbits in the reconstructed potential and compare them with ground-truth trajectories. %
    For an orbit initialized at the LMC center with the local circular velocity, the posterior-mean trajectory deviates from the true orbit by at most $1.0\,\mathrm{kpc}$ over $1\,\mathrm{Gyr}$, representing a substantial improvement over the $20.6\,\mathrm{kpc}$ deviation obtained with a near-truth analytic LMC potential (\autoref{fig:accel_orbit_errs}a).

    The reconstructed orbits exhibit stable energy evolution, with minimal fluctuations relative to both the true orbit and the initial inferred energy (\autoref{fig:accel_orbit_errs}b--c).
    In addition, the model accurately captures the structured, non-axisymmetric perturbation induced by the LMC.
    Fitting the full learned potential—combining the optimized analytic baseline with the neural-network residual—to a Milky Way + LMC parameterization recovers the LMC mass, scale radius, and Galactocentric distance to within $1.5\%$ of their true values, with all true values lying within $2\sigma$ of the BNN posterior median (\autoref{fig:accel_orbit_errs}d).

    \textbf{Time-dependent field reconstruction}:
    We extend the static model to a time‐dependent setting by modeling the LMC as a perturbation to a time‐evolving MW potential. %
    Training uses six time snapshots with 1024 samples each, and the evolution is learned in reverse from the present-day ($t=0$). %
    As shown in \autoref{fig:potential_evolution}, reconstruction errors remain under $1\%$ across a wide span of radii and epochs for both interpolation and extrapolation, with interpolation consistently more accurate. %

    \textbf{Network size: }
    \autoref{fig:depthwidth} summarizes how network depth and width affect training time and two performance metrics: the mean acceleration error (MAE), averaged over evaluation points, and the mean orbit deviation (MOD), averaged over a sample of test orbits. Physics-informed constraints keep architectures compact: even small networks achieve sub-percent MAE and sub-kiloparsec MOD, with training converging in approximately three minutes on a standard M2 CPU. %


\section{Discussion and future work}\label{sec:discussion}

    We show in this paper that embedding physics-informed constraints into the training process allows neural networks to reconstruct the acceleration and potential fields of galactic systems with high accuracy. %
    A critical next step is validation on realistic, noisy simulations to assess robustness and generalization beyond idealized test cases. %
    Moreover, because density fields are more directly observable than acceleration fields \cite{Silverwood_Easther_2019}, future extensions may benefit from adopting density-based loss functions, enabling more direct comparisons with observational data. %


\begin{ack} This work is supported by the National Science Foundation under Cooperative Agreement PHY-2019786 (The NSF AI Institute for Artificial Intelligence and Fundamental Interactions, http://iaifi.org/). \end{ack} 


\bibliographystyle{unsrtnat} \bibliography{refs} 
\appendix

\end{document}